\documentstyle[12pt]{article}
\input epsf
\newcommand{\be}{\begin{eqnarray}}
\newcommand{\ee}{\end{eqnarray}}

\begin{document}

\rightline{DESY98-198}
\rightline{RUB-TPII-20/98}

\begin{center}
{\large
Hard exclusive
pseudoscalar meson
electroproduction
and spin structure of a nucleon
}\\
\vspace{0.4cm}
{\bf
L.L.Frankfurt$^{a,b}$, P.V. Pobylitsa$^{b,c}$,
M.V. Polyakov$^{b,c}$, M. Strikman$^{b,d}$}\\
\vspace{0.3cm}
$^a$Physics Department, Tel Aviv University, Tel Aviv, Israel\\
$^b$Petersburg Nuclear Physics Institute, Gatchina, Russia\\
$^c$Institut f\"ur Theoretische Physik II, Ruhr--Universit\"at Bochum,\\
 D--44780 Bochum, Germany\\
$^d$Department of Physics, Pennsylvania
State University, University Park, PA 16802, USA,\\
and Deutsches Elektronen Synchrotron DESY, Germany\thanks{On leave of
absence from PSU.}\
\end{center}

\begin{abstract}
\noindent
The amplitude for  hard
exclusive
pseudoscalar meson electroproduction
off nucleon (nuclear) targets is computed in QCD within the
leading $\alpha_s \ln {Q^2/\lambda_{QCD}^{2}}$ approximation.
We show that the distribution of recoil nucleons depends strongly  on
the angle between the momentum of the recoil nucleon and the
 polarization vector of the target (or outgoing nucleon). This dependence
is especially sensitive to the spin flip skewed parton
distribution (SPD) $\widetilde E$.
We argue also that the
scaling for this spin asymmetry
sets in at lower $Q^2$ than
that for the absolute cross section.
Basing on the chiral quark-soliton model of the nucleon we estimate
quantitatively the spin asymmetry.
In the case of $\pi^+$ production this asymmetry is dominated
at small $t$ by the
contribution of the pion pole in the isovector SPD $\widetilde E$
as required by PCAC.
In the case of $K^0$ production off a proton
 we find a large enhancement of the  cross section as compared
to the case of $\pi^0$ production.
For the forward production of neutral pseudoscalar mesons off
a deuteron target we find  the cross section should be
zero for the zero deuteron helicity (along the
$\gamma^*D$ direction).
We consider also cross sections of quasielastic
processes off nuclei including the  feasibility to
implant
$K^+,\rho$-mesons into nuclear volume.
\end{abstract}

\section{Introduction}
\label{sec1}
Hard exclusive electroproduction of mesons is a new kind of
hard process calculable in QCD.
 In \cite{CFS} the QCD
factorization theorem was proven for the process
\begin{equation}
     \gamma ^{*}_L(q) + p \to  M(q-\Delta ) + B'(p+\Delta )
\label{process}
\end{equation}
at large $Q^{2}$, with $t$ and $x=Q^{2}/2p\cdot q$ fixed.
Here $M$ is any meson with   mass $m$
satisfying the condition:
$Q^2\gg m^2$. The factorization theorem asserts that
the amplitude has the form
\begin{eqnarray}
   &&
   \sum _{i,j} \int _{0}^{1}dz  \int d x_1
   f_{i/p}(x_1 ,x_1 -x;t,\mu ) \,
   H_{ij}(Q^{2}x_1/x,Q^{2},z,\mu )
   \, \phi _{j}(z,\mu )
\nonumber\\
&&
   + \mbox{power-suppressed corrections} ,
\label{factorization}
\end{eqnarray}
where $f_{i/p}$ is a skewed parton distribution (SPD)
\cite{Bartels,BB,Dittes,Abramowicz,CFS,Ji1,Rad,Jireview},
$x_1$ is the fraction of the  target momentum carried by
the  interacting parton,
$\phi_j$ is the light-front wave function of the
meson, and $H_{ij}$ is a hard-scattering coefficient,
computable in powers of $\alpha _{s}(Q)$.
Qualitatively one can say that these reactions allow
one to perform a
`` microsurgery'' of a nucleon by removing in a controlled way a
quark of one flavor and spin and implanting instead another quark
(in general with a different flavor and spin).
Use of the QCD factorization theorem makes it possible to predict
the  absolute
value of the cross section at sufficiently large $Q^2\ge$ 10GeV$^2$
and sufficiently small $x$ where skewed parton distributions
in a nucleon are calculable in QCD \cite{Freund,AMartin}.
For example, the exclusive neutral pion production is calculable
in terms of the valence quark distribution of nucleon spin
\cite{CFS,Piller,guichon}.
(An educated guess is that the region of applicability of the leading twist contribution to  hard
exclusive processes should be close to that for the electromagnetic
form factor of
the pion. Thus we expect that $Q^2\ge$ 10-15 GeV$^2$ should be a good
estimate, cf. discussion of the pion electromagnetic
form factors in \cite{Radyushkin,Krol}) and that approach
 to the scaling limit for pole contribution
would happen from above.

In this paper  we derive general formulae for the amplitude of
the hard exclusive
production of  pseudoscalar mesons off nucleons and nuclei and focus
on the calculation of the dependence of the  cross section
on the polarization of the target nucleon
which should be much less sensitive to
the inputs used in the calculation.
Really, an  experimental investigation of the spin asymmetries
discussed in the paper as well as ratios of
the yields of $\eta',\eta, \pi^0$
investigated in \cite{Eides}, and $K/\pi$ considered in this paper
has an important advantage. In fact, it was demonstrated in
\cite{FKS} that up to rather large $Q^2$ the absolute
cross sections of hard exclusive processes
are suppressed significantly due to the higher twist effects
originating from the  comparable transverse size of the longitudinal
photon and the meson wave functions. However the overall transverse
size was found to be sufficiently small $\le$ 0.4 fm at
 $Q^2 \ge$~5 GeV$^2$.
Due to the color transparency
phenomenon
this leads to  a strong suppression of the final state
interaction of the  $q \bar q$ pair which
in the end
will form the meson
and the residual baryon system. For $W \le 20$~GeV this cross section
is
of the order of few mb. Besides, the expansion of the
$q \bar q $ system
to a normal hadron size in the nucleon rest frame
takes a distance
$l_{coh} \sim 2p_M/\Delta m_M^2\sim
\frac{1}{x M_N} Q^2/\Delta m_M^2$ where $\Delta m_M^2$
is the characteristic light-cone energy denominator for a meson $M$
which is $\le$ 1 GeV$^2$, cf. \cite{fms}.
 One can easily see that the  condition $\l_{coh} \gg r_N$
is  satisfied for $x \le 0.2$ already for $Q^2 \ge$ 5 GeV$^2$.
Hence, it seems likely that
 {\it a precocious factorization into three blocks} -
overlap integral
between photon and pseudoscalar meson wave functions, the
hard blob, and the  skewed distribution
- could be valid already at moderately high
$Q^2$ leading to {\it a precocious scaling of the spin asymmetries and
of the ratios of the cross sections} as a function of  $Q^2$.
   At the same time predictions for the  absolute cross sections
should be valid only for  $Q^2 \geq 10$~GeV$^2$.

It is worth emphasizing also that the onset of the hard regime in the
case of  pion
electroproduction
(as well as the other channels for which
the vacuum exchange is forbidden) can be established
 via a study of the
change of the $t$ dependence
with an increase of $Q^2$.
For the production of electrically neutral
pseudoscalar mesons the $t$ dependence should asymptotically approach
with an increase of $Q^{2}$
to the value given by the
nucleon axial form factor.
 Besides due to
slowing down
of the Gribov diffusion in the hard regime one expects a significant
decrease of the shrinkage of the forward cone with the incident energy
at fixed $Q^2$. This effect would be easier to detect than in the
case of the vector meson production since the slope of the pion
trajectory is four times larger than that of  the Pomeron trajectory
\cite{FPScebaf}.

The paper is organized as following. In section \ref{sec2} we derive
general expressions for the process
of pseudoscalar meson production off nucleons. We relate the skewed
 proton $\to $ proton transitions to those arising for
 parton densities for the flavor changing transitions and give
predictions for the ratios of production of kaons and pions.
In particular we demonstrate that in contrast to the case for
soft physics where
production of strange particles is suppressed
in the hard exclusive processes strange meson production is
 often enhanced
as compared to the pion production. In particular we predict that for
small enough $x$ the cross section of the process
$\gamma^*_L +p \to K^0 +\Sigma^+$ should be three times larger than the
cross section for  the process $\gamma^*_L +p \to \pi^0 +p$.
In section \ref{sec21} we derive general expressions
for the  dependence of the meson production
cross section on the transverse proton polarization (or polarization
of the recoil baryon)
in terms of the skewed distributions.
In section \ref{sec3} we use the chiral quark soliton model of
 the nucleon \cite{pent} to calculate the relevant skewed distributions and
 to  calculate the polarization in the case of $\pi^+$ production,
and find it to be large due to the contribution of
the term controlled by PCAC.
In section \ref{nuclsec} we consider exclusive pion
production off nuclei. We find a number of spin effects
for these processes, and explain that color coherence phenomena
 should be
 present for these processes.
In section \ref{sec5} we introduce a new class of processes for
which the factorization theorem is applicable - exclusive production
of leading baryons and antibaryons. We discuss some general features
 of these
processes and suggest a way to use these processes to  look
 for gluonium states and to  implant mesons in nuclei.

We present the conclusions in section \ref{sec6}.

\section{Amplitude for the hard pseudoscalar meson electroproduction}
\label{sec2}
In this section we compute the leading $1/Q^2$
amplitude of the hard pseudoscalar meson electroproduction.
This amplitude owing to the factorization theorem for exclusive
hard reactions \cite{CFS} can be written as a convolution of
a skewed parton distribution in the nucleon, a
distribution amplitude  for  the
produced meson and the hard part computable in pQCD as
a  series in powers of
$\alpha_s$. We consider the processes of hard electroproduction of
a non-singlet pseudoscalar meson $P=\pi, K, $~etc.:
\be
\gamma_L^*(q) + B_1(p)\to P(q')+B_2(p+\Delta)\, ,
\ee
where $\Delta=p'-p$ is a four momentum transfer to a target
baryon $B_1$,
$-q^2=Q^2\to \infty$, $2 p\cdot q \to \infty$ with $x=Q^2/2 p\cdot q$
fixed.

The corresponding leading order amplitude has the form:

\be
\nonumber
&&\langle B_2(p'), P(p+q-p')| J^{\rm e.m.}\cdot
\varepsilon_L|B_1(p)\rangle=
-(e 4 \pi \alpha_s)\frac{N_c^2-1}{N_c^2} \frac{f_P}{4 Q}
\int_{-1}^1 d\tau \int_0^1 dz\\
&\times&\sum_{f,f'=u,d,s}
F^{(5)}_{ff'}(\tau,\xi,t) \varphi^{f'f}_P(z)
\Biggl\{ \frac{Q_{f'}}{z\bigl(\tau+\frac{\xi}{2} \bigr)-i0}+
\frac{Q_{f}}{(1-z)\bigl(\tau-\frac{\xi}{2} \bigr)+i0}
\Biggr\}\, .
\label{P-general}
\ee
Here $Q_f$ is the  charge of a quark of flavor $f=u,d,s$ in units of
the proton charge ($Q_u=2/3,\; Q_d=Q_s=-1/3$),
$f_P$ is the  decay constant of the pseudoscalar meson
($e.g.$ $f_\pi\approx 132$~MeV, $f_K\approx 1.2 f_\pi$, etc.), and
$\xi$ is a skewedness parameter (see definition below).

In the frame where $\vec{q}=(0,0,q_3)$ the longitudinal
polarization vector has the form:
\be
\varepsilon^\mu_L=
\frac{1}{Q} (q^3,0,0,q^0)^\mu
\, .
\label{polarization}
\ee
$F^{(5)}_{ff'}(\tau)$ is a skewed quark distribution
defined as:
\be
F^{(5)}_{ff'}(\tau,\xi,t)=
\int \frac{d\lambda }{2\pi }e^{i\lambda \tau}
\langle B_2(p^{\prime })|T \bigl\{
\bar \psi_{f'}
(-\lambda n/2){\hat n} \gamma_5
 \psi_f (\lambda n/2)\bigr\}
|B_1(p)\rangle  \; ,
\label{SPD1}
\ee
with the light cone vector $n$ normalized by:
\be
n^2 &=& 0,
 \hspace{2cm} n\cdot (p + p') \;\; = \;\; 2 \; .
\label{n-normalization}
\ee
The pseudoscalar meson distribution amplitude $\varphi^{f'f}_P(z)$
is defined as:

\be
\varphi^{f'f}_P(z)= \frac{1}{f_P}
\int \frac{d\lambda}{ 2\pi} e^{-i\lambda z (q'\cdot \widetilde n)}
\langle P(q')|T \bigl\{
\bar \psi_{f}
(\lambda \widetilde n){\hat{\widetilde n}} \gamma_5
 \psi_{f'} (0)\bigr\}
|0\rangle  \; ,
\label{DA}
\ee
where for convenience we introduce another light cone vector
$\widetilde n$ such that
$n\cdot \widetilde n=1$
and $\widetilde n^2 = 0$. Variable $z$ is the  longitudinal fraction
(along the vector $n$) of the meson momentum $q'=q+\Delta$ carried by one
of the quarks in a meson. Vectors $n$ and $\widetilde n$ are
linear combinations of the photon, initial, and final baryon momenta:
\be
\nonumber
p^\mu&=&(1+\frac \xi 2)\widetilde n^\mu+
(1-\frac \xi 2)\frac{\bar M^2}{2} n^\mu -\frac 12 \Delta_\perp^\mu\; \\
\nonumber
p^{\prime \mu}&=&(1-\frac \xi 2)\widetilde n^\mu+
(1+\frac \xi 2)\frac{\bar M^2}{2} n^\mu +\frac 12 \Delta_\perp^\mu\; \\
\nonumber
q^{\mu}&=&-\xi \widetilde n^\mu+
\frac{ Q^2}{2\xi} n^\mu \;\\
\bar M^2&=&\frac 12 (M_{B_1}^2+M_{B_2}^2-\frac{\Delta^2}{2})\, .
\label{kin}
\ee
The transverse plane is defined as the
 plane orthogonal to
the plane spanned by
 the light-cone vectors $n^\mu$ and $\widetilde n^\mu$.

The skewedness parameter $\xi$
(longitudinal component of the momentum transfer) is defined as
\begin{equation} \xi =-(n\cdot \Delta ).
\label{xi-def} \end{equation}
In the Bjorken limit when $Q^2\to \infty$ but $x$ is fixed this
skewedness parameter $\xi$ is expressed in terms of the Bjorken variable
$\xi=\frac{2 x}{2-x}$.

In the expression (\ref{P-general}) $a~ new$ type of
``truly non-diagonal" skewed parton distributions (SPD's) enter;
they are given by the transitional matrix element $B_1\to B_2$
of a bilocal (generically non-diagonal in the flavor space) quark operator
on the light-cone (\ref{SPD1}). One can use the flavor $SU(3)$
or $SU(2)$ symmetry to relate these distributions to the usual
SPD's corresponding to the diagonal transition
proton (neutron)$\to$proton (neutron), which were
introduced to describe the deeply virtual Compton scattering (DVCS)
amplitude.
One can write for a quark-antiquark operator on a light-cone
between two baryons
\begin{equation}
\langle B_2|\bar q_iq_j|B_1\rangle =F[B_2^{+},B_1]_{ji}+D\{B_2^{+},B_1\}_{ji}
+ S \delta_{ji}\delta_{B_2B_1}
\end{equation}
Since we are interested in the
nondiagonal transitions we can ignore the singlet
part. Taking the matrix elements of nonsinglet combinations of
$\bar q_iq_i$ operators between the proton states we find:
\begin{equation}
\langle p|\left( \bar uu-\bar dd\right)=D-F
|p\rangle
\end{equation}
\begin{equation}
\langle p|\left( 2\bar uu-\bar dd-\bar ss\right) |p\rangle =2(D-F)-(D+F)=D-3F
\end{equation}
\begin{equation}
\langle p|\left( \bar dd-\bar ss\right) |p\rangle =-(D+F)
\label{su3}
\end{equation}

Hence  in the case of the  production of $\pi^+$, which is described
in terms of the transitional matrix element between proton and neutron
we have (the spin indices are suppressed):
\be
\langle n| \bar d u|p\rangle=
\langle p| \bar u u|p\rangle-
\langle n| \bar u u|n\rangle=
\langle p| \bar u u|p\rangle-
\langle p| \bar d d|p\rangle\; ,
\label{isospin}
\ee
to relate these $new$ SPD's to the usual ones.
The relation (\ref{isospin})
and $SU_{fl}(3)$ relations below should be understood as flavor relations
between matrix elements (\ref{SPD1}). They do not imply any relations
between, for example,
 valence and sea quark distributions. For instance, relation
(\ref{isospin}) states for fixed $\tau$ that:
\be
F^{(5) (p\to n)}_{ud}(\tau)=
F^{(5) (p\to p)}_{uu}(\tau)-
F^{(5) (p\to p)}_{dd}(\tau) \, .
\ee
Another example is the  production of kaons.
For instance,
$K^+$ production probes
$p\to\Lambda$ or $p\to\Sigma^0$ transitions in
eq.~(\ref{P-general}). In this case eq.(\ref{su3})
leads to
(only flavor quantum numbers are shown)
\footnote{
Let us note that although we expect that $SU_{fl}(3)$ relations work
rather well for spin nonflip SPD's $\widetilde H$, they can be violated
rather strongly for ``the pole part" of
$\widetilde E$ (see discussion in section~4).
This is due to  the rather large mass difference $m_K-m_\pi$.
However the ``pole part" is under theoretical control,
because it is calculated using PCAC.}:
\be
\nonumber
\langle \Lambda | \bar s u |p\rangle=\frac 1{\sqrt{6}}(3F-D) &=&
\frac{1}{\sqrt{6}}
\langle p| (2 \bar u u -\bar d d -\bar s s)|p\rangle\, ,\\
\langle \Sigma^0 | \bar s u |p\rangle= \frac 1{\sqrt{2}}(F+D)&=&
\frac{1}{\sqrt 2}
\langle p|
\bar d d- \bar s s|p\rangle\; .
\label{su32}
\ee
In the  case of $K^0$ production  from a  proton we probe the
$p\to\Sigma^+$ transition in eq.~(\ref{P-general}). The $SU_{fl}(3)$
relation in this case is  the following:
\be
\langle \Sigma^+ | \bar s d |p\rangle=(F+D) &=&
\langle p|
\bar d d- \bar s s|p\rangle\; .
\label{su31}
\ee

We see that the hard electroproduction of pseudoscalar mesons
gives the  possibility to probe various flavor combinations of
polarized quark distributions in the proton
and to study in a new way $SU_{fl}(3)$ properties of the strange baryons.

Analogously we can write the flavor decomposition of the
pseudoscalar meson distribution amplitudes (\ref{DA}).
Here we give some examples for pions and kaons:
\be
\nonumber
\varphi^{f'f}_{\pi^0}(z)&=&
\frac{i}{\sqrt 2} \bigl[
\delta_u^{f'}\delta_u^{f} -
\delta_d^{f}\delta_d^{f'}
\bigr]
\varphi_\pi(z) \, ,\\
\nonumber
\varphi^{f'f}_{\pi^+}(z)&=&
i\,
\delta_d^{f'}\delta_u^{f}
\varphi_\pi(z) \, ,\\
\nonumber
\varphi^{f'f}_{K^0}(z)&=&
i\,  \delta_s^{f'}\delta_d^{f}
\varphi_K(z) \, ,\\
\varphi^{f'f}_{K^+}(z)&=&
i\,
\delta_s^{f'}\delta_u^{f}
\varphi_K(z) \, ,
\label{DA2}
\ee
where $\varphi_\pi(z)$ and $\varphi_K(z)$ are the leading twist
pion and kaon distribution amplitudes normalized by the condition:
$\int_0^1 dz \varphi_{\pi,K}(z)=1$.
In principle, other components of the light cone
meson wave function may give a nonzero contribution,
but only in subleading order in $1/Q^2$. Thus we took into account
in the  leading order only the conventional wave function of
a  pion and a kaon
given by the  matrix element of the axial current.

The leading twist skewed quark distribution
with $f=f'=q=u,d,s$ and $B_1=B_2=$proton
given by eq.~(\ref{SPD1})
can be decomposed into spin nonflip and spin flip parts.
Here we adopted the notation of Ji \cite{Ji1} for the spin decomposition
of the matrix element of a bilocal quark operator between protons.
The spin nonflip part is denoted as $\widetilde H_q$ and
spin flip as $\widetilde E_q$. They are defined as:

\be
\nonumber
\int \frac{d\lambda }{2\pi }e^{i\lambda \tau}\langle p^{\prime }|\bar
\psi_q
(-\lambda n/2){\hat n} \gamma_5
 \psi_q (\lambda n/2)|p\rangle
&=& \widetilde H_q(\tau,\xi ,t) \;
\bar U(p^{\prime }) \; \hat n \gamma_5 \; U(p) \\
&+& \frac 1{2M_N} \; \widetilde E_q(\tau,\xi ,t) \;
\bar U(p^{\prime }) \; (n\cdot  \Delta) \gamma_5 U(p) .
\nonumber \\
\label{E-H-QCD-2}
\ee
Here
$\Delta$ is  the four--momentum transfer, $\Delta = p^{\prime }-p$,
$M_N$ denotes the nucleon mass, and $U(p),\bar U(p^{\prime })$
are the  standard Dirac spinors.
One can chose as independent variables for the
skewed quark distributions, $\widetilde H_q(\tau,\xi ,\Delta ^2)$
and $\widetilde E_q(\tau,\xi ,\Delta ^2)$, the variable
$\tau$, related to the fraction of target momentum
carried by the  interacting parton $x_1$ by:
\be
\tau=\frac{x_1-\frac{x}{2}}{1-\frac{x}{2}}\;
\ee
 the square of the four--momentum transfer, $\Delta^2=t$, and
the light-cone fraction of $\Delta$ - $\xi$ (the skewedness parameter)
\footnote{
The skewed parton distributions
and meson distribution amplitudes are scale dependent;
the scale is set by the photon
virtuality $Q^2$. We do not show the scale dependence of
these quantities so as  to simplify the  notation.}.

In the forward case, $p = p'$, both $\Delta$ and $\xi$ are zero, and
the second term on the r.h.s.\ of eq.(\ref{E-H-QCD-2}) disappears. The
function $\widetilde H_q$ becomes the usual polarized parton distribution
function,
\be
\widetilde H_q(\tau, \xi = 0, t= 0) &=& \Delta q(\tau)
=\left\{
\begin{array}{cr}
\Delta q(\tau)=q_+(\tau)-q_-(\tau),& \hspace{.5cm} \tau \; > \; 0\,, \\
\Delta \bar q(-\tau)=\bar q_+(-\tau)-\bar q_-(-\tau)
,& \hspace{.5cm} \tau \; < \; 0 \,.
\end{array}
\right.
\label{forward_limit}
\ee
Here $q_\pm(\tau)$ are densities of partons with positive (negative)
helicity in the proton with positive longitudinal polarization.
Let us note that the universal function
$\widetilde H_q(\tau, \xi = 0, t= 0)$
corresponds to the sum of valence and sea polarized quark distributions.
At the same time only
the valence polarized quark distributions
give nonzero contribution to the  amplitudes of
the electroproduction of
neutral
pseudoscalar mesons.
We want to draw attention that these distributions
are not constrained by inclusive DIS experiments.
This is because of the  negative charge parity
in the $t-$channel for
the amplitude for electroproduction of electrically neutral
pseudoscalar meson.
However these distributions are probed in semiinclusive
DIS experiments.

It is useful to define isoscalar:
\be
\nonumber
\widetilde H^{(0)}(\tau,\xi,t)&=&
\widetilde H_u(\tau,\xi,t)+
\widetilde H_d(\tau,\xi,t) \, ,\\
\widetilde E^{(0)}(\tau,\xi,t)&=&
\widetilde E_u(\tau,\xi,t)+
\widetilde E_d(\tau,\xi,t) \, ,
\ee
and isovector
\be
\nonumber
\widetilde H^{(3)}(\tau,\xi,t)&=&
\widetilde H_u(\tau,\xi,t)-
\widetilde H_d(\tau,\xi,t) \, ,\\
\widetilde E^{(3)}(\tau,\xi,t)&=&
\widetilde E_u(\tau,\xi,t)-
\widetilde E_d(\tau,\xi,t) \, ,
\ee
skewed quark distributions.

The SPD's $\widetilde H^{(3)}(\tau,\xi,t)$ and
$\widetilde E^{(3)}(\tau,\xi,t)$ satisfy the sum rules \cite{Ji1}:
\be
\nonumber
\int_{-1}^1 d\tau \widetilde H^{(3)}(\tau,\xi,t)= G_A(t)\, ,\\
\int_{-1}^1 d\tau \widetilde E^{(3)}(\tau,\xi,t)= G_P(t)\, ,
\ee
where $G_A(t)$ ($G_A(0)=g_A\approx 1.25$) and $G_P(t)$ are the axial
and pseudoscalar form factors of the nucleon.

In the  next sections we will consider
polarization effects in the production of pseudoscalar mesons off a
polarized proton. To be specific we will
discuss pion electroproduction.
 All formulae can be easily generalized
to the case of production of other pseudoscalar mesons
using the general expression for the production amplitude (\ref{P-general})
and isospin ($SU_{fl}(3)$) relations (\ref{isospin},\ref{su32}).

For the charged pion production the leading order amplitude
can be written as:
\be
\nonumber
&&\langle n(p'), \pi^+(p+q-p')| J^{\rm e.m.}\cdot
\varepsilon_L|p(p)\rangle=
-(ie 4 \pi \alpha_s)\frac{N_c^2-1}{N_c^2} \frac{1}{24 Q}
f_\pi \int_0^1 dz \frac{\varphi_\pi(z)}{z}\\
\nonumber
&&\times\Biggl\{ \bar U(p')\hat n \gamma_5 U(p)
\int_{-1}^1 d\tau
\widetilde H^{(3)}(\tau,\xi,t)\Bigl(3 \alpha^-(\tau)-\alpha^+(\tau)\Bigl) \\
&&+\bar U(p')\frac{ n\cdot\Delta}{2 M_N} \gamma_5 U(p)
\int_{-1}^1 d\tau
\widetilde E^{(3)}(\tau,\xi,t)\Bigl(3 \alpha^-(\tau)-\alpha^+(\tau)\Bigl)
\Biggl\}\, ,
\label{ampp}
\ee
where
\be
\alpha^{\pm}(\tau)=\frac{1}{\tau+\frac{\xi}{2}-i0}\pm
                \frac{1}{\tau-\frac{\xi}{2}+i0}\, ,
\ee
is the hard kernel computed in the leading order of $\alpha_s$.

The neutral pion production amplitude has the form
(this amplitude was obtained previously in
\cite{Piller,guichon}\footnote{Ref.~\cite{Piller}used a
different  spin decomposition of SPD's from those here.}):
\be
\nonumber
&&\langle p(p'), \pi^0(p+q-p')|J^{\rm e.m.}\cdot
\varepsilon_L|p(p)\rangle=
-(ie 4 \pi \alpha_s)\frac{N_c^2-1}{N_c^2}
\frac{1}{24 \sqrt{2}  Q}
f_\pi \int_0^1 dz \frac{\varphi_\pi(z)}{z}\\
\nonumber
&&\times\Biggl\{ \bar U(p')\hat n \gamma_5 U(p)
\int_{-1}^1 d\tau
\Bigl(\widetilde H^{(3)}(\tau,\xi,t)+3\widetilde H^{(0)}(\tau,\xi,t)
\Bigl)\alpha^+(\tau)\\
&&+\bar U(p')\frac{ n\cdot\Delta}{2 M_N} \gamma_5 U(p)
\int_{-1}^1 d\tau
\Bigl(\widetilde E^{(3)}(\tau,\xi,t)+3\widetilde E^{(0)}(\tau,\xi,t)
\Bigl)\alpha^+(\tau)
\Biggl\}\, .
\label{amp0}
\ee

Let us illustrate at the end of this section the possible applications
of the $SU_{fl}(3)$ relations eqs.~(\ref{isospin},\ref{su32},\ref{su31}).
With their help we can roughly estimate
the ratios of yields of kaons to pions in hard exclusive
 electroproduction
reactions.
For instance, we can easily estimate the ratio of yields of $\pi^+(n)$
and $K^+(\Lambda)$ at small $x$ ($x< m_\pi/M_N$)
\footnote{For larger $x$ we have to take into account
large contributions
of $\pi^{\pm}$ and $K^+,K^0$ poles to the spin flip SPD $\widetilde E$
(see discussion below).} as:
\be
 \frac{K^+}{\pi^+}
\approx \frac{f_K^2}{6\ f_\pi^2}\frac{
[3(2\Delta u_s-\Delta d_s -\Delta s_s)
-(2\Delta u_v-\Delta d_v -\Delta s_v)]^2}{
 [3(\Delta u_s - \Delta d_s) -(\Delta u_v - \Delta d_v)]^2} \, .
\ee
Here we introduce valence $\Delta q_v=\Delta q-\Delta \bar q$
and sea $\Delta q_s=\Delta q+\Delta \bar q$
 polarized quark distributions.
These quantities are
 antisymmetric and symmetric under transposition
$s \leftrightarrow u$ and have signatures -1 and 1.
(Note that our definition of $\Delta q_s$ differes from the often used
definition  $\Delta q_s= 2\Delta \bar q$.)
 We want to draw attention
 that in this case
both the  sea quark ($\bf{8}_F$) and valence quarks($\bf{8}_D$)
 contribute to the amplitude, see eq.(\ref{su32}).

For the ratio of $K^0$ and $\pi^0$ yields
we get the estimate:
\be
 K^0:\pi^0 \approx 2 f_K^2 (\Delta d_v -\Delta s_v)^2 :
  f_\pi^2 (2 \Delta u_v + \Delta d_v)^2 \approx 3:1\, ,
\ee
where we take  for simplicity $\Delta s_v =\Delta u_v+\Delta d_v\approx 0$
for $x=0.1\div 0.4$ (this is also in line with large $N_c$
counting, see below).
Surprisingly large ratios!

\section{Spin asymmetry}
\label{sec21}
The expressions for the amplitudes (\ref{ampp},\ref{amp0})
contain the
full information about the reaction $\gamma^*_L +p\to
\pi^a+N$ in the leading order of $1/Q^2$ expansion.
In the present paper we consider a specific polarization observable, for
which we expect (see the  discussion in the introduction)
a precocious scaling at
moderately large
$Q^2\sim 2\div 4 $ GeV$^2$.
We expect that for the absolute
cross section the onset of  scaling occurs at larger values of
$Q^2\ge 10$~GeV$^2$. Also the spin asymmetry is the most sensitive
observable to probe the spin-flip SPD $\widetilde E$.

Let us consider production of the pion off a polarized proton.
The differential cross section can be written in the form:

\be
\sigma=\sigma_0+\sigma_1
([ \vec p^\prime_\perp,\vec S_\perp]\cdot \vec e_z)/|\vec p^\prime_\perp|=
\sigma_0+\sigma_1 |\vec S_\perp| \sin\beta \; ,
\ee
where  $S_\perp$ is the transverse
 projection of the polarization vector of the initial proton
(or of the polarization vector of the outgoing nucleon
taken with ``-'' sign),
$\vec e_z$ is a vector   normal to this plane and $\beta$
is the  angle between the $S_\perp$ and $p^\prime_\perp$ (transverse
component of the momentum of the outgoing nucleon).

Using the above expressions
for the leading twist amplitudes of the
charged (\ref{ampp}) and neutral (\ref{amp0}) pion production
we define:
\be
\nonumber
A_+&=& \int_{-1}^1 d\tau
\widetilde H^{(3)}(\tau,\xi,t)\Bigl(3 \alpha^-(\tau)-\alpha^+(\tau)\Bigl)\\
B_+&=& \int_{-1}^1 d\tau
\widetilde E^{(3)}(\tau,\xi,t)\Bigl(3 \alpha^-(\tau)-\alpha^+(\tau)\Bigl)\, ,
\label{ABp}
\ee
for the charged pion production, and
\be
\nonumber
A_0&=& \int_{-1}^1 d\tau
\Bigl(\widetilde H^{(3)}(\tau,\xi,t)+3\widetilde H^{(0)}(\tau,\xi,t) \Bigl)\alpha^+(\tau)\\
B_0&=& \int_{-1}^1 d\tau
\Bigl(\widetilde E^{(3)}(\tau,\xi,t)+3\widetilde E^{(0)}(\tau,\xi,t) \Bigl)\alpha^+(\tau)\, ,
\label{AB0}
\ee
for the neutral pion production.
For definitness let us consider the
following asymmetry:
\be
{\cal A}=\frac{1}{|S_\perp|}
\frac{
\int_0^{\pi}d\beta |{\cal M}(\beta)|^2-
\int_{\pi}^{2\pi}d\beta |{\cal M}(\beta)|^2
}{\int_{0}^{2\pi}d\beta |{\cal M}(\beta)|^2}=
\frac{2\sigma_1}{\pi \sigma_0}\, .
\ee
It follows from
our definition of the kinematical variables (\ref{kin}) that
$p'_\perp=\Delta_\perp/2$ with
$\Delta_\perp^2=
-\frac{4(1-x)}{(2-x)^2}(t-t_{\rm min})
+O(1/Q^2)$, where
$t_{\rm min}=-M_N^2 x^2/(1-x)+O(1/Q^2)$ is the minimal
(in  absolute value) momentum transfer squared in the reaction.
In terms of SPD's the asymmetry ${\cal A}$ has the form
(for the  charged and neutral pion production):
\be
{\cal A}_{+,0}=
\frac{|\Delta_\perp|}{\pi M_N}
\frac{\xi {\rm Im}(A_{+,0} B_{+,0}^*)}{
|A_{+,0}|^2 (1-\frac{\xi^2}{4})-|B_{+,0}|^2
\frac{t \xi^2}{16 M_N^2}-\frac{\xi^2}{2}{\rm Re}(A_{+,0}B_{+,0}^*)}\, .
\label{asym}
\ee
{}From this expression we see immediately that the asymmetry ${\cal A}$
would be zero if the spin flip SPD $\widetilde E$ were zero.
In contrast to the SPD $\widetilde H_q(\tau,\xi,t)$ which in the forward limit
$t\to 0$  reduces to the usual polarized quark distributions
$\widetilde H(\tau,\xi=0,t=0)=\Delta q(\tau)$,
the SPD $\widetilde E(\tau,\xi,t)$
can not be related to already known quark distributions.
For the estimates of the asymmetry (\ref{asym}) we shall resort to
the computation of the
skewed parton distributions in the large $N_c$ limit in
the chiral quark-soliton model \cite{pppbgw,pent}.

\section{Skewed parton distributions in the large $N_c$ limit
and spin asymmetry}
\label{sec3}
Recently the skewed quark distribution $\widetilde H$ and $\widetilde E$
were computed in the chiral quark-soliton model of the nucleon \cite{pent}.
In this model nucleons can
be viewed as $N_c$ ``valence" quarks bound by a self-consistent pion
field (the ``soliton") whose energy coincides with the aggregate
energy of the quarks of the negative-energy Dirac continuum. Similarly to
the Skyrme model,
the large $N_c$ limit
is needed
to justify the use of
the mean-field approximation.

In \cite{DPPPW} a new approach to the calculation of quark distribution
functions
has been developed within the context of the chiral
quark-soliton model of the nucleon \cite{DPP}, and furthermore  in \cite{pppbgw}
this method was applied to calculate SPD's.
The range of the applicability of the chiral quark-soliton model to the
parton distributions is limited by the conditions:
$\left|t\right|\ll M_N^2$,
$\tau,\xi \sim 1/N_c$, the computed distributions refer to
a  low normalization point of about $600$~MeV.

According to the large $N_c$ analysis of ref.~\cite{pent}
$\widetilde H^{(0)}\ll \widetilde H^{(3)}$ and
$\widetilde E^{(0)}\ll \widetilde E^{(3)}$
in the large $N_c$ limit
which  is in line with the phenomenology
for the diagonal case.
Another crucial observation made in \cite{pent} is that the
isovector skewed quark distribution $\widetilde E^{(3)}$ has
a strong enhancement at small $t$
originating from the long range pion tail of the nucleon
wave function.

This contribution has the form \cite{pent}:

\be
\widetilde E^{(3)}_\pi=\theta\left[|\tau|<\frac{|\xi|}{2}\right]
\Phi_\pi\Bigl(\frac{2\tau}{\xi}\Bigl)
\frac{2F(t)}{\xi}\, ,
\label{pionpoler}
\ee
where $\Phi_\pi(z)$ is the pion distribution amplitude, normalized
by $\int_{-1}^1 dz \Phi_\pi(z)=1$
($\Phi_\pi(z)=\frac 12 \varphi_\pi(\frac{1-z}{2})$).
The pion distribution amplitude
$\Phi_\pi(x)$ in eq.~(\ref{pionpole}) calculated in the same model
\cite{pp,pp2} is very close to the asymptotic one
$\Phi_\pi(z)=\frac34 (1-z^2)$, so we shall always use the asymptotic
pion distribution amplitude for numerical calculations.
The form factor $F(t)$ in this model is related to the Fourier
transform of the mean pion field $U(x)=\exp(i \pi^a(x) \tau^a)$:

\be
F(-\vec k^2)= \frac{2 M_N^2 f_\pi^2}{3k^3}\int d^3 x\ \exp(i\vec k
\cdot \vec x) {\rm Tr}[(U(x)-1)\tau^3]\, .
\ee
Taking into account that at large distances the soliton field
has the Yukawa tail (see e.g. \cite{DPP}):
\be
\lim_{r\to \infty }(U(x)-1)=\frac{i x^a\tau^a}{r}
\frac{3 g_A}{4\pi f_\pi^2 r^2}(1+m_\pi r)\exp(-m_\pi r)\; ,
\ee
one gets immediately the small $t$
asymptote
of the form factor
$F(t)$:
\be
\lim_{t\to m_\pi^2} F(t)=
-\frac{4 g_A M_N^2}{t-m_\pi^2}\, .
\label{pionpole}
\ee
The numerical results for the form factor $F(t)$
can be found in \cite{pent}.
Let us note that the form factor
$F(t)$ in a  wide range of $t$: $m_\pi^2\ll |t| \ll M_N^2$
differs significantly from the pole contribution
(\ref{pionpole}).
For example, for $-t$=0.35~GeV$^2$ they differ by a factor of two.
In our numerical estimates of the
asymmetry we shall use the results of \cite{pent}.

The pion pole contribution (\ref{pionpole})
corresponds to the pion pole contribution
to the nucleon form factor $G_P(t)
$\footnote{A similar expression is valid
for the contribution of the kaon pole.}:
\be
\lim_{t\to m_\pi^2}
\int_{-1}^1 d\tau \widetilde E^{(3)}_\pi(\tau,\xi,t) =
-\frac{4 g_A M_N^2}{t-m_\pi^2}\, ,
\ee
required by spontaneously broken chiral symmetry. We see
that the appearance of the pion pole in $\widetilde E^{(3)}$ is a
general consequence of PCAC and in order to reproduce it in some model
it should respect the
chiral Ward identities.
For example, in the computation of SPD's in the bag model \cite{Ji3}
the chiral Ward identities are violated and the pion pole contribution
(\ref{pionpole}) is missed.
The chiral quark-soliton
model respects all chiral Ward identities that
 allows one to split unambiguously
SPD $\widetilde E^{(3)}$ into two pieces \cite{pent}:
\be
\widetilde E^{(3)}(\tau,\xi,t)=\widetilde E^{(3)}_\pi(\tau,\xi,t)+
\widetilde E^{(3)}_{\rm smooth}(\tau,\xi,t)\, ,
\ee
the result for $\widetilde E^{(3)}_{\pi}$ is given by eq.~(\ref{pionpoler});
the results for $\widetilde E^{(3)}_{\rm smooth}(\tau,\xi,t)$ can be found
in \cite{pent}.

It is easy to see that $\widetilde E^{(3)}_\pi(\tau,\xi,t)$ does not
contribute to the $B_0$ amplitude of the neutral pion production
(see eqs.~(\ref{AB0})). However, it contributes to the $B_+$ amplitude
of the charged pion production (see eqs.~(\ref{ABp})). The corresponding
contribution has the form:

\be
B_+^{(\pi)}=\frac{12 F(t)}{\xi}
\int_{-1}^1 dz \frac{\Phi_\pi(z)}{1-z^2}\,.
\ee
We see that the SPD $\widetilde E^{(3)}_\pi$ eq.(\ref{pionpoler})
contributes only to the real part of
amplitude $B_+$, also it is strongly enhanced at small $t$ and $\xi$.
Using the results of ref.~\cite{pent} we obtain that
Re$B_+^{\pi}$$\gg$Re$B_+^{\rm smooth}$,Im$B_+$
at least by an order of
magnitude for $|t|<0.5$~GeV$^2$ and $0.1<x<0.4$.
In the
 case of the neutral pion production with the spin flip amplitude
$B_0$, the pion pole does not
show up.
As a result $B_0$ and $B_+^{\rm smooth}$ should have the same magnitude.
This result immediately implies that

\be
{\cal A}_0 \ll {\cal A}_+\, .
\ee

Neglecting Re$B_+^{\rm smooth}$ and Im$B_+$ relative to
$B_+^{(\pi)}$ we get the following expression for the asymmetry
(\ref{asym}) ${\cal A}_+$ of charged pion production

\be
{\cal A}_+\approx
\frac{|\Delta_\perp|}{\pi M_N}
\frac{18 \eta F(t)\ {\rm Im}A_{+}}{
|A_{+}|^2 (1-\frac{\xi^2}{4})-
\frac{81\eta^2 t}{4M_N^2} F(t)^2
-9\xi F(t){\rm Re}A_{+}}\, ,
\label{asympi}
\ee
where $\eta = \frac 23 \int_{-1}^1 dz \frac{\Phi_\pi(z)}{1-z^2}$
($\eta=1$ for the asymptotic pion distribution amplitude).

In Fig.~1 we plot the asymmetry (\ref{asympi}) at several values
of $t$ as a function of $x$. We plot the asymmetry
for  $\eta=1$ (corresponding to the asymptotic pion distribution
amplitude). We show also the result for  $\eta=5/3$
(corresponding to Chernyak-Zhitnitsky
model for the pion DA)
merely to illustrate
the sensitivity of the asymmetry to
the shape of the pion distribution amplitude. The instanton model gives the
almost asymptotic pion distribution amplitude (see \cite{pp,pp2})
already at a low normalization point of about $600$~MeV.
One can see that the spin
asymmetry in the production of charged pions is of order unity, this
makes challenging the measurements of this asymmetry.
Alternatively, one can measure the recoil nucleon polarization.
 Also it is seen that the
 sensitivity to the form of the pion distribution
amplitude is maximal at small $t$.

Let us analyze eq.~(\ref{asympi}) at small momentum
transfer $|t|< m_\pi^2$ and small Bjorken $x < m_\pi/M_N$.
In this range of $t$ and $x$
the contribution of the pion pole (the spin flip part
of the amplitude) to the
absolute cross section is suppressed, whereas
for the spin asymmetry we have the following expression:
\be
\nonumber
{\cal A}_+&\approx&
\frac{72 |\Delta_\perp| \eta g_A M_N}{\pi m_\pi^2}
\frac{{\rm Im}A_{+}}{|A_{+}|^2} \\
&=&
\frac{144 |\Delta_\perp| \eta g_A M_N}{ m_\pi^2}
\frac{\biggl(2
\widetilde H^{(3)}(\frac \xi 2,\xi,t)
+\widetilde H^{(3)}(-\frac \xi 2, \xi,t)\biggr)}{|A_{+}|^2}
\, .
\label{asympichir1}
\ee
We see that the asymmetry in
the electroproduction of
charged pions
in this limit depends on $both$
the polarized valence and sea
quark distributions. The particular combination which enters in this
case is poorly
constrained by the DIS measurements.
This makes the hard exclusive
pion production experiments a useful source of information on
the polarized quark distributions.\footnote{Note that if the process
$e+p \to e+ \pi^+ +n $ is studied via a setup detecting only the scattered
electron and $\pi^+$ one may not have good enough energy resolution
to separate
 this process from the process $e+p \to e+ \pi^+ + \Delta^0$.
However our estimates using the methods
outlined in \cite{FPScebaf} indicate
that this would result only in a slight
dilution of the asymmetry.}

\section{Hard Coherent production of pions off nuclei}
\label{nuclsec}

In this section we discuss briefly
the hard exclusive production of pseudoscalar mesons off
nuclei. We focus  primarily
on the polarization effects and comment also on
the  peculiarities
of the
$A$-dependence related to the spin structure of the
elementary amplitude.

First let us
consider the scattering off  deuteron targets.
We explained above that the amplitude for the electroproduction of
a neutral pseudoscalar meson (\ref{amp0}) (in contrast to the charged
pion production amplitude) is dominated at small $t$
by the  helicity
non-flip amplitude $A_0$.
In the (near) forward limit
$A_0$
is
proportional to the spin of
 nucleon -
$\phi(p')\sigma_z\phi(p),$ where $z$ is the photon momentum direction
and $\phi(p)$ is the spin part of the wave function of
the  nucleon.
Therefore, in the case of the
coherent
scattering off a polarized deuteron target at small $t$
the dominant
term is proportional to the projection of
the sum of the operators of the nucleon spins
to the reaction axis
with the proportionality coefficient
calculable in terms of
the isoscalar polarized valence quark distributions.
If this projection is zero, the corresponding amplitude
should be zero for the forward scattering. This
is because
the matrix element of the projection of the sum of
spins of nucleons is equal to zero
even if the $D$-wave in the
deuteron wave function is taken into account.
In the case of the charged pion production dominance of the spin-flip
amplitude  leads to a strong excitation of the isotriplet two-nucleon
virtual state.

Meson production off
heavier nuclei in the discussed limit
constitutes  another class  of  color coherent phenomena.
Since the mesons are produced in  small size configurations they do not
interact with the residual system
within the leading twist approximation.
Hence the amplitudes of such processes
are  expressed through the skewed parton density of
the nucleus and can be represented (neglecting small EMC like effects)
as a convolution of the  elementary amplitude and
the nuclear transition amplitude:
\be
M(\Delta)= \int d^3\vec{k} \left<A'(\vec{\Delta})\right|\delta(H_{A'}-\Delta_0)
a^{\dagger}_{s',N_f}(\vec{k}+\vec{\Delta})
a_{s,N_i}(\vec{k})\left|A\right>,
\ee
where $\Delta$ is the four-momentum transfer in the nucleus rest frame, and
$a^{\dagger},a$ are production and annihilation operators for nucleons
with given spin states. A correlation between these spin states is
 implicitly
contained in the elementary amplitude. Such matrix elements can
be calculated
using conventional
methods of   low-energy nuclear physics.
For large enough $t$ one can use closure
over the processes of nuclear disintegration
$A'$
 and write (we give the answer for production of neutral mesons):
\be
\sum_{A'} {d\sigma(\gamma^*_L+A \to M +A')\over dt}=
Z{d\sigma(\gamma^*_L+p \to M +p')\over dt}+
N{d\sigma(\gamma^*_L+n \to M +n')\over dt},
\ee
where $Z$ and $N$ are the  numbers of protons and neutrons in the
target.

For small enough $t$  the result is strongly
sensitive to the spin effects. Let us give several examples.

(a) For the case of $\pi^0$ production off
  $^3He$ one finds  for small $t$
a strong cancelation of the contribution of the scattering off the protons
(in the approximation that $^3He$ is described as a $S$-wave system).
\be
{d\sigma(\gamma^*_L+^3He \to \pi^0 +^3He)\over dt}=
{d\sigma(\gamma^*_L+n \to \pi^0 +n)\over dt}F_{A=3}(t),
\ee
where $F_{A=3}(t)$ is a superposition of the magnetic and electric
 $^3He$ form factors
normalized to one at $t=0$.

(b) for reaction $\gamma^*_L+ ^3He \to \pi^+ +^3H$ for small $t$
 one expects spin asymmetries
similar to the case of scattering off a proton which we discussed above.
If,  in line with our expectations,
the spin-flip amplitude dominates in the elementary cross section
for small $t$,  the
cross section will be approximately equal to
\be
{d\sigma(\gamma^*_L+^3He \to \pi^+ +^3H)\over dt}=
{d\sigma(\gamma^*_L+p \to \pi^+ +n)\over dt}F_{M, A=3}(t),
\ee
where $F_{M, A=3}(t)$ is the magnetic form factor of $^3He$
normalized to one at $t=0$.

(c) In the case of $\pi^0$ production
off $^4He$ for small $t$ we expect a  much
stronger suppression of the cross section than in the case of
scattering off the deuteron. Indeed, in this case
the dominant contribution in the elementary
amplitude is a vector in the spin
indices which cancels out when averaged with the
$\left|in\right>$ and $\left<out\right|$ spin zero states of
$^4He$.

The $K^+$ meson production is
 also of interact for the study of the low-energy
hyperon-nucleus interactions. In a sense,
 we have here a situation of a local
``micro" implantation of a hyperon with a small momentum in
any point in the nucleus, predominantly in the central region,
without disturbing any other nucleons. This is markedly different from
the pion induced processes where for heavy nuclei only production near
the surface is important while
 the final state interactions of the
outgoing kaon
cannot be neglected.

To summarise,
studies of the scattering off nuclei will provide  new handles
for the study of  spin effects,  effects of color coherence,
 as well as of
the spin structure of low-energy interactions.

\section{Exclusive production of forward baryons off nucleons
and meson implantation in nuclei}
\label{sec5}
The proof of the  factorization for the meson exclusive production,
eq.~(\ref{P-general}),
is essentially based on the observation that the
  cancellation of the soft gluon interactions
is
intimately related to the fact that
the meson arises from a
quark-antiquark pair generated by the hard scattering.  Thus the
pair starts as a small-size configuration and only substantially
later grows to a normal hadronic size, to a meson.  This
implies that the parton density is a standard parton density
(apart from the skewed  nature of its definition).
Therefore 
the  factorization theorem is valid
also for  the  production of leading baryons
\begin{equation}
     \gamma ^{*}(q) + p \to  B(q+\Delta ) + M(p-\Delta )
\label{process1}
\end{equation}
and even leading antibaryons
\begin{equation}
     \gamma ^{*}(q) + p \to  \bar{B}(q+\Delta ) + B_2(p-\Delta ),
\label{process2}
\end{equation}
where $B_2$ is a system with the baryon charge of two.
Processes (\ref{process1},\ref{process2}) will provide 
 unique information
about multiparton correlations in nucleons.
For example,  process (\ref{process1}) will allow one to
investigate 
the probability 
for three quarks in a nucleon to
come close together without collapsing the wave function into
a three quark component
 - 
 such a probability would not be small
in meson cloud models of the nucleon and in, say, the  MIT bag model.
 On the other hand if one would try
to follow an analogy with the case of the 
positronium,  this probability would be strongly suppressed.
At the same time   reaction (\ref{process2})
would allow one to determine a 
probability (presumably numerically very small) to have
 three antiquarks close together in a nucleon.

To describe in QCD  process
(\ref{process1}),  a new
 non-perturbative mathematical
object should be introduced
 in addition to the usual
distribution amplitudes (DA).
It can be called a skewed
distribution amplitude (SDA.)
It is
 defined as
a non-diagonal matrix element of the tri-local quark operator
between a meson $M$ and a proton:
\be
\nonumber
&&\int \prod_{i=1}^3 dz_i^- \exp[i\sum_{i=1}^3 x_i\, (p\cdot z_i)]\,
\langle M(p-\Delta)|\varepsilon_{abc}
\, \psi^{a}_{j_1}(z_1) \,
\psi^{b}_{j_2}(z_2) \,
\psi^{c}_{j_3}(z_3)|N(p)\rangle\Bigl|_{z_i^+=z_i^\perp=0}\\
&&=\delta(1-\zeta-x_1-x_2-x_3)\, F_{j_1 j_2 j_3}(x_1,x_2,x_3,\zeta,t) \;,
\ee
where $a,b,c$ are color indices, $j_i$ are spin-flavor indices, and 
$F_{j_1 j_2 j_3}(x_1,x_2,x_3,\zeta,t)$ are the  new  SDA.
These SDA
can be decomposed into
invariant spin-flavor structures
which depend on the quantum numbers of the meson $M$.
They
depend on the variables $x_i$ (which are contracted with
the  hard kernel in
the amplitude), on the skewedness parameter $\zeta=1-\Delta^+/p^+$
(in some sense, with this definition of $\zeta$ the limit
$\zeta\to 0$ corresponds to the usual distribution amplitude, $i.e.$
skewedness $\to 0$ means SDA $\to$ DA)
and the momentum transfer squared $t=\Delta^2$.

Though quantitative
calculations of  processes (\ref{process1}, \ref{process2})
 will take  time, some
qualitative predictions could be made
right away: the cross section
of the process for fixed $x$, and large $Q^2$
 should be proportional to the baryon
elastic form factor. In particular, it would be instructive to study
the $Q^2$ dependence of the ratio of  the  cross section for the
 process $\gamma ^{*}(q) + p \to p +
\pi^0$ and the square of the elastic proton form factor.
If the color transparency suppresses the final state interaction
between the fast moving nucleon and the residual meson state
 early enough
one may expect that this ratio may
reach the scaling limit in the region where
higher twist contributions to the nucleon form factor are still large.
(It is worth emphasizing that the longitudinal
distances involved in the final state interaction of the system flying
along $\vec q$ with the residual system are
 much smaller in this case than in the case of the 
$A(e,e'p)$ reaction, so the expansion
effects would be much less important in this case.)
Another  interesting process is $\gamma^{*}(q) + p \to \Delta^{++} +
\pi^-$  which may allow one to compare the wave functions of 
the $\Delta$-isobar
and a nucleon in  a way
complementary to the $N\to \Delta$ transition processes.
Note also, that
a 
study of the meson spectrum in the process $\gamma^* +p \to p +M$
may help to investigate the role of pions in the nucleon wave function.
In a naive model of the nucleon with
a pion cloud one would expect that the single pion 
production will dominate.
However if more complicated nonlinear pion  fields are important in
the low $Q^2$ $q \bar q$ sea, one may expect that
 higher
recoil
masses
 would be at least as important.

Reaction (\ref{process1}) provides also 
a promising avenue
to look for exotic meson states including gluonium.
Indeed, if one would consider, for example, the MIT bag model, 
the  removal of
 three quarks from the system  could leave the
residual system looking asw like a bag made predominantly  of glue.
It is natural to expect that such a system would have a large overlapping
integral with gluonium states.

An interesting property of  reaction (\ref{process1}) is that
it could be used to produce a  meson ``$M$" ($\eta, \rho, K$-meson
 with  zero momentum in the nucleon target rest frame) provided
\begin{equation}
q_0={Q^2+2m_{M}m_N-m_{M}^2 \over 2(m_N-m_{M})}.
\end{equation}
In the case of the 
production of a leading baryon off the nuclear target,
similar to the case of
the  production of leading mesons discussed
in section \ref{nuclsec},
the outgoing nucleon would
not interact with the nucleus since it is produced in a
small-size configuration.
Hence it is possible in the discussed reactions to ``implant"
a meson
 with a small momentum in the center of the nucleus.
 This implantation
 has a number of advantages as compared to the processes of
low-energy implantation since in
 the low energy processes  mesons are produced
 predominantly near the surface and the interaction of other
hadrons involved in the process is, in general, large.
Therefore process (\ref{process1}) could be used for
studying the medium modification of the properties of various mesons.
In the case of  $\rho$-meson 
production an additional advantage
of this process as compared to low-energy
processes is that in the  high-energy case the  effects
 of distortion of the  two pion mass
spectrum due threshold effects are much less important.

\section{Conclusions}
\label{sec6}
We have derived the 
general expression for the hard pseudoscalar meson
($\pi, K$, etc.) electroproduction amplitude in QCD in 
the leading
order in $1/Q^2$ and $\alpha_s$.

We have demonstrated that the
distribution of the recoil nucleons in the hard exclusive production
of $\pi^+$ strongly depends on
the angle between the momentum of the
recoil nucleon and the spin polarization
of the target (or the outgoing nucleon). This dependence is
especially sensitive
to the helicity flip skewed parton distribution (SPD) $\widetilde E$.
This helicity flip SPD contains at small momentum transfer
the contribution of the pion pole due to the chiral Ward identities.
Owing to the factorization theorem for the hard exclusive processes
\cite{CFS} the relative size of pole and non-pole contributions can be
estimated quantitatively using the {\it low-energy}
chiral model of the nucleon
\cite{pent}.

For $x\rightarrow 0$ (i.e. $t=t_{min}\rightarrow 0$)   hard
electroproduction of charged
mesons gives a 
unique possibility to measure the $x$ dependence of the polarized valence
quark distribution in 
a nucleon (nuclear) target. This is because
skewed parton distributions are calculable in this limit
in terms of the conventional valence quark distributions in a target.
Such an investigation  is 
especially interesting since in this limit this $x$ dependence
 should be different from the one for the unpolarized case
due to different 
 quantum numbers in the crossed
channel.
Additionally, comparing reactions 
with different flavor
quantum numbers, like $e.g.$:
\be
\nonumber
&&\gamma_L^* +p \to \pi^+ +n \, ,\\
\nonumber
&&\gamma_L^* +p \to K^+ +\Lambda \, ,\\
\nonumber
&&\gamma_L^* +p \to K^+ +\Sigma^0 \, ,\\
\nonumber
&&\gamma_L^* +p \to K^0 +\Sigma^+ \, ,
\ee
one can access different flavor combinations of polarized quark
distributions in the proton,
 and probe
$SU_{fl}(3)$ for spin part of the strange baryon quark densities
(see eqs.~(\ref{isospin},\ref{su32},\ref{su31})).

We have also demonstrated that the study of exclusive meson production off 
the nuclear target will provide addition information 
about the spin structure
of skewed distributions as 
well as allow one  to look for color coherence effects.

\section{Acknowledgments}
We would like to acknowledge discussions with
K.~Goeke, L.~Mankiewicz, G.~Piller, A.~Radyuskin, A.~Shuvaev,
M.~Vanderhaeghen and C.~Weiss.
We thank I.~B\"ornig and M.~Penttinen for help with numerical
calculations. 
We thank J.Whitmore for reading the paper and useful comments.
The work of P.V.P. and M.V.P.
has been supported  by RFBR grant 96-15-96764 and the
 joint grant of RFBR and the Deutsche
Forschungsgemeinschaft (DFG) 436 RUS 113/181/0 (R), by the BMBF grant
RUS-658-97 and the COSY (J\"ulich).
M.S. would like to thank DESY for hospitality
during the time this work was done. The work of 
M.S. and L.F.  is supported in part by the U.S. Department of
Energy and BSF.

\noindent
After completing the paper we learned that the
cross section of $\pi^+$ hard electroproduction
was
calculated independently
by L.~Mankiewicz, G.~Piller and A.~Radyushkin \cite{MPR}.
We are grateful to them for the discussions
and an exchange of results.

\newpage
\begin{figure}
\setlength{\epsfxsize}{15cm}
\setlength{\epsfysize}{15cm}
\epsffile{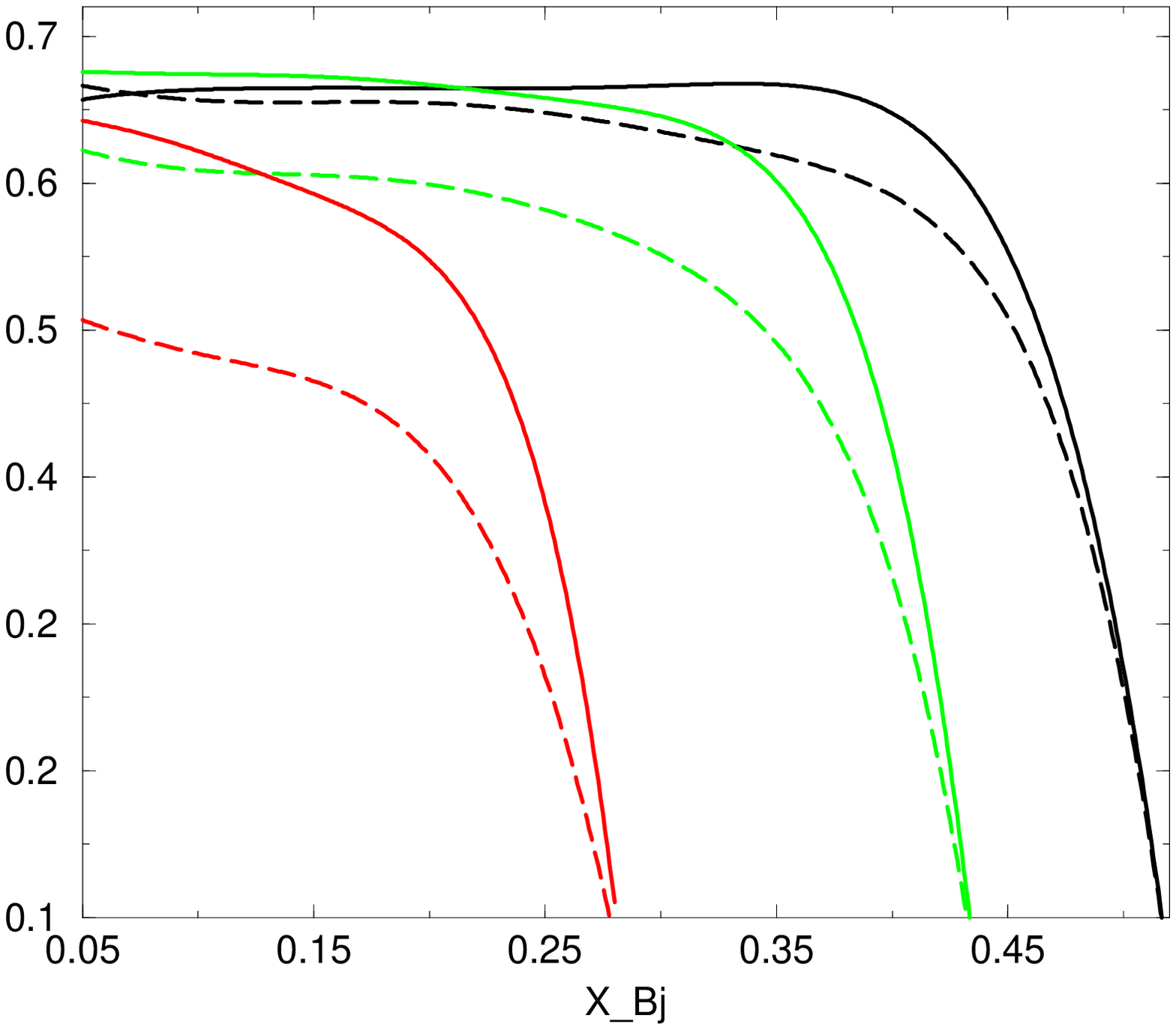}
\caption[]
{The asymmetry ${\cal A}_+$ as a function of Bjorken $x$
at $t=-0.1$, -0.3, -0.5 GeV$^2$.
{\em Solid lines}: With asymptotic pion distribution amplitude
$\eta=1$
{\em Dashed lines:} With
Chernyak-Zhitnitsky pion distribution amplitude $\eta=5/3$}
\label{fig_fig4}
\end{figure}

\end{document}